# Cosmic Ray Physics with ACORDE at LHC

C. Pagliarone[1]*, A. Fernandez-Tellez[2]
[1] Università degli Studi di Cassino & INFN Pisa,  Largo B. Pontecorvo, 3 - Pisa, Italy.
[2] Benemérita Universidad Autónoma de Puebla (BUAP), Puebla, Mexico.

E-mail: pagliarone@fnal.gov

**Abstract.** The use of large underground high-energy physics experiments, for comic ray studies, have been used, in the past, at CERN, in order to measure, precisely, the inclusive cosmic ray flux in the energy range from $2 \cdot 10^{10} \div 2 \cdot 10^{12}$ eV. ACORDE, ALICE Cosmic Rays DEtector, will act as Level 0 cosmic ray trigger and, together with other ALICE apparatus, will provide precise information on cosmic rays with primary energies around $10^{15} \div 10^{17}$ eV. This paper reviews the main detector features, the present status, commissioning and integration with other apparatus. Finally, we discuss the ACORDE-ALICE cosmic ray physics program.

## 1. Introduction

The study of Cosmic-Ray (CR) physics, using high energy collider detectors, have been explored, achieving important results, by L3 and ALEPH collaborations at CERN Large Electron Positron collider (LEP) [1]. The need, for ALICE, to have a CR trigger (CRT), not only for calibration and alignment purposes, brought us to further explore the possibility of widening the ALICE physics program by contributing with a genuine CR physics program.

## 2. The ACORDE detector

ACORDE, ALICE Cosmic Ray DEtector is an array of plastic scintillator modules (60 at the present), placed on the top 3 sides of the central ALICE magnet, as shown in Figure 1. More modules, to achieve a better angular coverage and acceptance, may be added later on. Each ACORDE module, consists of two plastic scintillator paddles with $190 \times 19.5$ cm$^2$ effective area, arranged in a doublet configuration as shown in Figure 2. Each doublet consists of two superimposed scintillator counters, with their corresponding photomultipliers active faces, looking back to back. A coincidence signal, in a time window of 40 ns, coming from the two scintillator paddles, gives, for each module, the CRT trigger hit. A PCI BUS electronics card have been developed in order to measure plateau and efficiency of the module counters [2]. The CRT will provide a fast level-zero trigger signal to the central trigger processor, when an atmospheric µ impinge upon the ALICE detector. The signal will be useful for calibration, alignment and performance studies of other ALICE tracking detectors such as: the Time Projection Chamber (TPC), the Transition Radiation  Detector (TRD) and the Inner Tracking System (ITS)[3]. A typical rate, for a single atmospheric µ, crossing the ALICE cavern, is ≤ 1 Hz/m$^2$. The expected rate, for multi-µ events, is ≤ 0.04 Hz/m$^2$. Only atmospheric µ with energy larger than 17 GeV  will reach ACORDE;  the TPC will be able, to reconstruct  muons with energy up to  2 TeV. The

---
* Corresponding author and the person to whom any correspondence should be addressed.

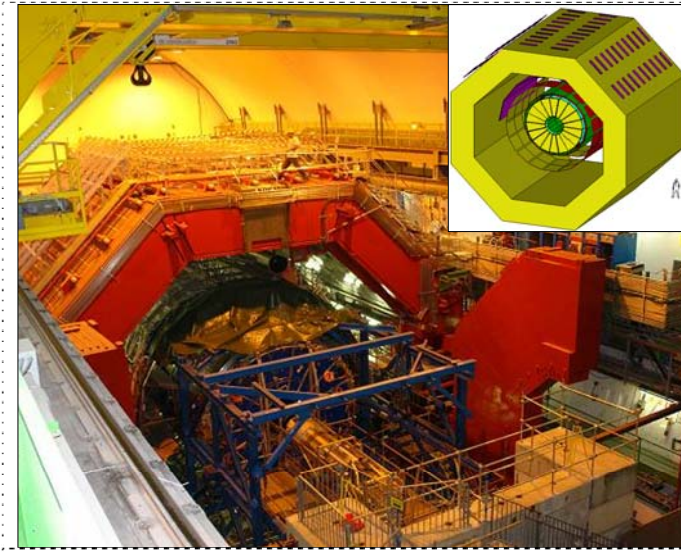 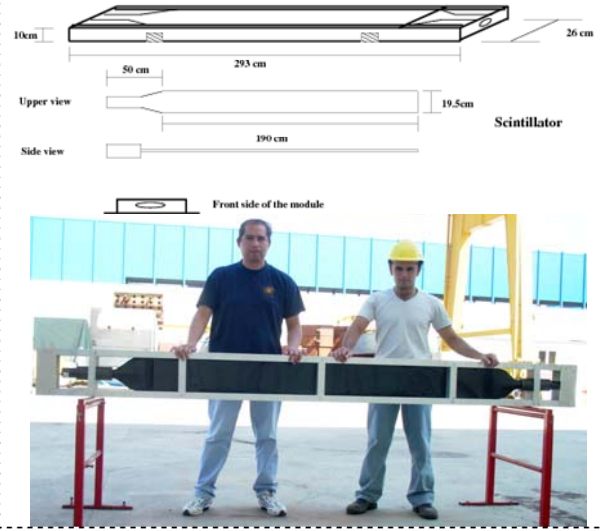

**Figure 1.** The ALICE Cosmic Ray Detector; the scintillator array is on the three top faces of the magnet.

**Figure 2.** A schematic view of an ACORDE module and a picture of one ready to be installed.

The ACORDE electronics is essentially made of 60 Front End Electronics cards (FEE), one for each module; an ACORDE OR card used to generate the TRD wake up signal (this card receive the 60 coincidence LVDs signals coming from the FEE cards); a Main Card, which contains all the electronics needed to receive the 120 LVDS signals coming from the scintillator counters (this card produce single and multi-coincidence trigger signals and provide the connectivity to the ALICE trigger and DAQ systems) [4]. At the present, the ACORDE commissioning is proceeding well. We have 20 modules working and taking data. The remaining 40 are installed and cabled. The electronics is presently under integration with the DAQ and CTP systems and the detector control system is ready.

### 3. CR Physics with ACORDE

There are two main CR physics aspects where ALICE could contribute. The first would lead to a better understanding of nuclear interactions at very high energy. This aspect is fundamental in order to describe the CR interactions in atmosphere. The particle production, both at large energies and in the forward direction, can be estimated only using models based on the extrapolation of accelerator data. The interpretation of cosmic ray data, in particular the identification of primary cosmic rays, rely crucially on such a models. In those models, the calculations of $\sigma_{p-Air}$ leads to different results that start to diverge at energies relevant for CR physics at the knee $(3 \cdot 10^{15} \cdot eV)$. Presently, most of the interaction models, such as VENUS, QGSJET, DPMJET, neXus, SIBYLL, HDPM, etc., rely on *pp*, *ep* and heavy-ion collisions data. On the other hand, at present, the highest energy data comes from Tevatron (1.96 TeV). ALICE, with is unique design, mainly developed to track and identify a very large number of charged particles, created in pp, pA and *AA* collisions, may be the an ideal place where to measure such cross sections. The second aspect is a direct contribution to CR physics accessible by using these detector peculiarities together with ACORDE CRT. The underground location of ALICE, with *30 m* of overburden composed of sub-alpine Molasses, is an ideal place for a muon-based underground experiment. The average energy threshold, for muons reaching the central magnet, after crossing the rock, is $E_{\mu th} \geq 17\ GeV$. Energy loss and multiple coulombian scattering change the properties of muons reaching the experiment, affecting both the flux, the shape and the energy spectrum. From a geological surveys, made at the L3 time, we know that the surface, above the ALICE location, is flat within a radius of at least *200 m* from the ALICE interaction point and the rock composition is almost uniform. We are able, then, to measure accurately the μ absorption and energy loss (see Figure 3). ALICE is particularly well adapted to observe underground multi-muon events, which are of interest in order to investigate the composition of very high energy CRs and to search for rare exotic CR events. Our physical goals are mainly to search for large muon multiplicities (≈100) in multi-muon events. The transverse size of muon bundles, their number and their energy distributions are sensitive to the mass of the primary nucleus inducing the Extended Air Shower (EAS). Possible exotic scenarios related with this subject are collisions of strange quark matter (SQM) with air nuclei or formation of quark-gluon-plasma (QGP) in Fe-air collisions. We are also planning to measure the vertical muon flux and spectrum. Available results, coming from other experiments, show discrepancies of up to 15–20% in the absolute normalization of the muon fluxes and differences in the observed energy spectra. A new measurement using ALICE with the ACORDE CRT, on vertical muons, would clarify this situation, making use of the powerful tracking capabilities and the high statistics that the experiment can produce in several years of data taking. Other possible searches, such as search for point sources or the measurement of the antiproton/proton ratio, at very high energies, maybe be in principle possible [5]. A proposal of adding surface detectors, spread over a large distances at CERN, to search for correlations between multi-muon events and large area showers, is under consideration.

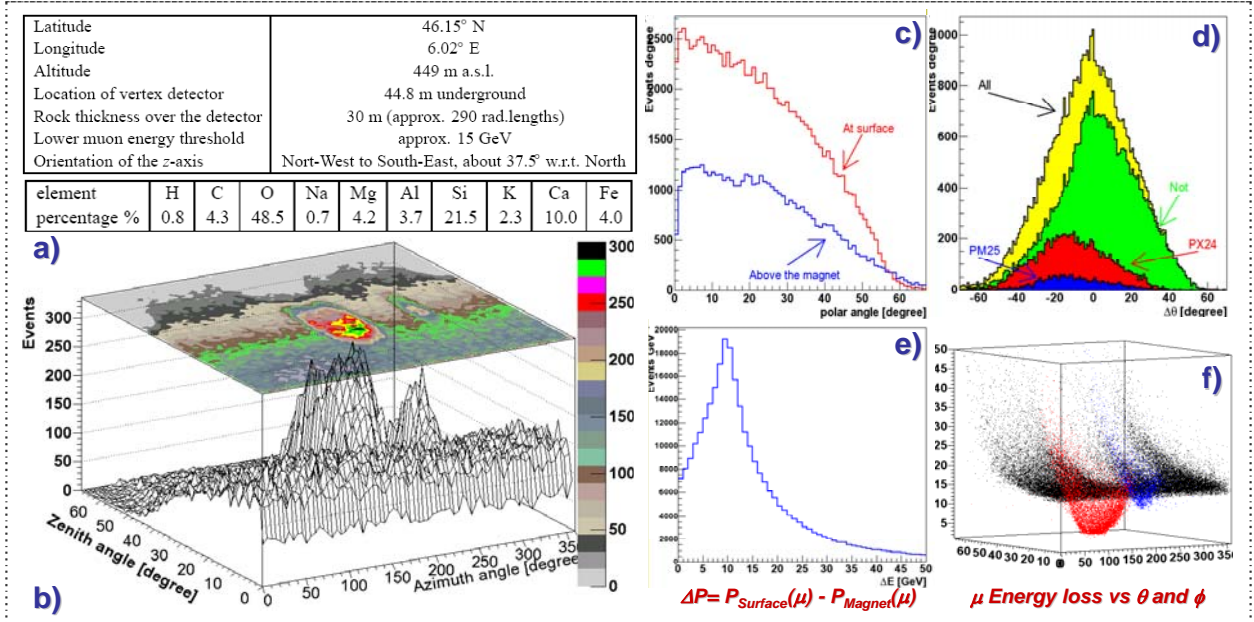

**Figure 3. a)** ALICE environment and rock composition; **b)**, **f)** μ energy loss versus θ and φ; **c)** μ polar distribution, **d)** μ polar deflection, **e)** μ energy spread, at the CERN ground and on the ALICE magnet.

## 4. Acknowledgements

This work was supported by CONACyT (Project No. 47318/A-1) and by the ALFA-EC funds in the framework of the Program HELEN (High Energy Physics Latinoamerican-European Network).